\documentclass[prl, aps, reprint, amsfonts, amssymb, amsmath, preprintnumbers, showpacs, nofootinbib, superscriptaddress]{revtex4-1}

\usepackage{graphicx, multirow,soul}
\usepackage[citecolor=blue]{hyperref}
\usepackage[all]{hypcap}
\usepackage{url}
\usepackage{color}
\usepackage{subfigure}
\usepackage{slashed}
\usepackage{float}
\usepackage{bbm}
\usepackage{bbold}
\usepackage{ulem}
\usepackage{lipsum}

\newcommand{\figref}[1]{Fig.~\ref{#1}}

\newcommand{\be}{\begin{equation}}
\newcommand{\ee}{\end{equation}}
\newcommand{\ba}{\begin{aligned}}
\newcommand{\ea}{\end{aligned}}
\usepackage{tikz}
\tikzset{node distance=2cm, auto}
 \usepackage[compact]{titlesec}
 \titlespacing{\section}{0pt}{2ex}{1ex}
 \titlespacing{\subsection}{0pt}{1ex}{0ex}
  \titlespacing{\subsubsection}{0pt}{0.5ex}{0ex}

\newcommand{\GeV}{{\rm ~GeV}}

\newcommand{\g}{{\rm ~g}}
 \newcommand{\ie}{i.e.}

\hypersetup{colorlinks,citecolor= nicegreen,linkcolor= nicered}
\definecolor{nicered}{rgb}{0.7,0.1,0.1}
\definecolor{nicegreen}{rgb}{0.1,0.5,0.1}
\definecolor{nicepurple}{RGB}{0, 76, 151}
\definecolor{ypgcm}{RGB}{0, 102, 51}

\allowdisplaybreaks

\begin{document}

\title{Assessing the tension between a black hole dominated early universe and leptogenesis}

\author{Yuber F.~Perez-Gonzalez}
\email{yfperezg@northwestern.edu; https://orcid.org/0000-0002-2020-7223}
\affiliation{Theoretical Physics Department, Fermi National Accelerator Laboratory, P.O. Box 500, Batavia, IL 60510, USA}
\affiliation{Department of Physics $\&$ Astronomy, Northwestern University, Evanston, IL 60208, USA}
\affiliation{Colegio de Fisica Fundamental e Interdisciplinaria de las Americas (COFI), 254 Norzagaray
street, San Juan, Puerto Rico 00901}

\author{Jessica Turner}
\email{jessica.turner@durham.ac.uk; https://orcid.org/0000-0002-9679-5252}
\affiliation{Institute for Particle Physics Phenomenology, Department of
Physics, Durham University, South Road, Durham DH1 3LE, United Kingdom.}

\date{\today}

\begin{abstract}
We perform the first numerical calculation of the interplay between thermal and black hole induced leptogenesis, demonstrating that the right-handed neutrino surplus produced during the evaporation only partially mitigates the entropy dilution suffered by the thermal component. As such, the intermediate-mass regime of the right-handed neutrinos, $10^6\GeV \lesssim M_{N} \lesssim 10^{9}\GeV$, could not explain the observed baryon asymmetry even for fine-tuned scenarios if there existed a primordial black hole dominated era, consistent with initial black hole masses of $M_i \gtrsim  \mathcal{O}\left(1\right)$ kg. Detection of the gravitational waves emitted from the same primordial black holes would place intermediate-scale thermal leptogenesis under tension.
\end{abstract}

\preprint{FERMILAB-PUB-20-528-T} 
\preprint{NUHEP-TH/20-10} 
\preprint{IPPP/20/46} 
\maketitle
%%%%%%%%%%%%%%%%%%%%%%%%%%%%%%%%%%%%%%%%%%%%%%%%%%%%%%%%%%
%%%%%%                          Introduction                             													
%%%%%%%%%%%%%%%%%%%%%%%%%%%%%%%%%%%%%%%%%%%%%%%%%%%%%%%%%%
\textit{ Introduction.} --- 
Primordial black holes (PBHs) could have been formed shortly after the Big Bang through many different mechanisms~\cite{Carr:2020gox,Carr:2020xqk,Khlopov:2008qy}. Once created,  PBHs constitute an increasingly large portion of the universe's energy budget and play an essential role in the universe's evolution.

PBHs were initially proposed by Hawking who explored their quantum mechanical properties \cite{Hawking:1974rv,Hawking:1974sw}. This work led to his discovery that black holes evaporate over time. The fate of these PBHs is not known: it is possible they completely disappear or leave a relic which can constitute a portion of dark matter \cite{Chapline:1975ojl,MacGibbon:1987my,Carr:2016drx,Ali-Haimoud:2016mbv,Carr:2009jm,Bird:2016dcv,Inomata:2017okj}. Nonetheless,
during their evaporation, the PBHs will produce all possible particles with masses below the temperature of the black hole. 
This democratic feature of PBHs can lead to the production of dark matter \cite{Morrison:2018xla,Fujita:2014hha,Lennon:2017tqq}, dark radiation \cite{Hooper:2019gtx,Lunardini:2019zob,Hooper:2020evu,Masina:2020xhk,Keith:2020jww} and  the  observed matter antimatter asymmetry. 
The latter idea was initially explored in \cite{Hawking:1974rv,Carr:1976zz,1976Caot,Baumann:2007yr}
where heavy, new particles, produced from PBH evaporation, can decay in a CP- and baryon-number- 
violating manner to produce the observed baryon asymmetry \cite{Patrignani:2016xqp,Ade:2015xua}.

Several works examine the connection between neutrino masses,
the matter antimatter asymmetry and PBHs \cite{Fujita:2014hha,Morrison:2018xla}.
These works explore a scenario where dark matter and right-handed neutrinos
are produced via Hawking radiation. The subsequent non-thermal decays of the right-handed neutrinos
violate CP and lepton number, resulting in the generation of a 
 lepton asymmetry converted to a net excess of baryons via non-perturbative Standard Model effects \cite{Khlebnikov:1988sr}. 
 This process, known as \textit{leptogenesis}, links the origin of light neutrino masses, suppressed by the heavy, right-handed neutrino
mass, with the baryon asymmetry. 

In this \textit{Letter}, we compute, for the first time,
the numerical solutions to the Friedmann equations for the evolution of a PBH dominated universe
and the baryon asymmetry produced from thermal leptogenesis together with the non-thermal contribution generated by the PBH evaporation.
We show that the baryon asymmetry undergoes a dilution due to the entropy injection from the black hole evaporation.
We quantify such depletion in the intermediate-scale regime, $10^6\lesssim M_{N} (\text{GeV}) \lesssim 10^9$, and find that  even for the most optimistic scenarios in this regime, the baryon asymmetry would be insufficient to explain the observed value for initial PBH masses $M_i \gtrsim  \mathcal{O}\left(1\right)$ kg. Finally, we discuss the gravitational wave spectrum and highlight that such signals' future detection would falsify thermal leptogenesis in the intermediate regime. Our code will be made publicly available through the {\tt ULYSSES} python package~\cite{Granelli:2020pim}.
We will consider natural units throughout.

%%%%%%%%%%%%%%%%%%%%%%%%%%%%%%%%%%%%%%%%%%%%%%%%%%%%%%%%%%%
%%%%%%                                          SeeSaw                                				%%%%%%
%%%%%%%%%%%%%%%%%%%%%%%%%%%%%%%%%%%%%%%%%%%%%%%%%%%%%%%%%%%
\textit{Type I Seesaw Mechanism.} --- 
The seesaw mechanism \cite{Mohapatra:1979ia,GellMann:1980vs,Yanagida:1979as,Minkowski:1977sc,Mohapatra:1980yp,Magg:1980ut,Lazarides:1980nt,Wetterich:1981bx} is a compelling model that can simultaneously explain  the smallness of neutrino masses and the origin of the matter antimatter asymmetry. In the most minimal framework, the Standard Model (SM) Lagrangian is augmented to include at least two heavy right-handed neutrinos with masses $M_{N_i}$:
\be
\mathcal{L} = i\overline{N_{i}}\slashed{\partial}N_{i}  -\overline{L_{\alpha}}Y_{\alpha i}N_{i}\tilde{\Phi}-\frac{1}{2}\overline{N^C_{i}}M_{N_i}N_{i} + \text{h.c.}\,,
\ee
where $N_i$, $L_{\alpha}$ and $\Phi$ denote the right-handed neutrinos of generation $i$, $\rm{SU}\left(2\right)_{L}$ leptonic doublets of flavor $\alpha$ and Higgs doublets, respectively, with the  negative hypercharge Higgs doublet  defined as $\tilde{\Phi} = i\sigma_{2}\Phi^*$. In its original manifestation \cite{Fukugita:1986hr}, the right-handed neutrinos can decay out of thermal equilibrium at temperatures similar to their mass. Further, if the Yukawa matrix, $Y_{\alpha i}$, contains complex phases these decays can produce more anti-leptons than leptons (or vice versa) and the resultant lepton asymmetry  is converted via electroweak sphaleron processes  \cite{Khlebnikov:1988sr} to a baryon asymmetry. This mechanism is commonly known as non-resonant thermal leptogenesis and the minimal mass scale is $M_{N_i}\gtrsim 10^{6}$ GeV \cite{Moffat:2018wke}\footnote{In this regime, the mass splittings of the right-handed neutrinos is order one and therefore far from the resonant regime. This was also explored in \cite{Moffat:2018smo}.}.

The mass of the right-handed neutrino can be lowered to the TeV regime if right-handed neutrinos are highly degenerate in mass \cite{Pilaftsis:1997jf,Pilaftsis:2003gt} or lower still, to the GeV scale, if leptogenesis via oscillations is the production mechanism of the baryon asymmetry~\cite{Akhmedov:1998qx,Asaka:2005pn}.  We focus on minimal non-resonant thermal leptogenesis and its interplay with a black hole dominated earlier universe.

The Yukawa matrix can be conveniently parametrized such that neutrino oscillation data is automatically recovered \cite{Casas:2001sr},
\be\label{eq:Yuk}
Y = \frac{1}{v}U\sqrt{m_\nu}R^T\sqrt{M_{N}}\,,
\ee
where $v= 174 \text{ GeV}$, $U$ is the leptonic mixing matrix, $m_{\nu}$ is the diagonal light neutrino mass matrix, $R$ is a complex, orthogonal matrix and $M_N$ is the diagonal mass matrix of the heavy right-handed neutrinos. The model parameter space is 18 dimensional where nine parameters are associated with the low-energy scale physics, and the remaining nine parameters are associated with the high-scale physics of the right-handed neutrinos.

%%%%%%%%%%%%%%%%%%%%%%%%%%%%%%%%%%%%%%%%%%%%%%%%%%%%%%%%%%%%%%%%%%%%%%%%%%%%
%\section{RIGHT HANDED NEUTRINO PRODUCTION FROM PRIMORDIAL BLACK HOLES}\label{sec:right-handedNfromPBH}
%%%%%%%%%%%%%%%%%%%%%%%%%%%%%%%%%%%%%%%%%%%%%%%%%%%%%%%%%%%%%%%%%%%%%%%%%%%%
\textit{Right-handed neutrino production from Hawking radiation.} --- 
Black holes (BHs) evaporate by emitting a thermal flux of particles, known as Hawking radiation \cite{Hawking:1974rv,Hawking:1974sw}. Such a flux is generated because of the gravitational disruption created by the collapsing matter forming the BH \cite{birrell_davies_1982}. The emission is democratic in nature, \ie, all particles existing in the universe can be emitted independently of their interactions. Therefore, if right-handed neutrinos do indeed exist, they would be among the particles producing during the evaporation. The instantaneous emission rate of right-handed neutrino $N_i$, with momentum between $p$ and $p+dp$ and time interval $dt$, is \cite{Hawking:1974rv,Hawking:1974sw}
\begin{align}\label{eq:nuHS}
	\frac{d^2 {\cal N}_{N_i}}{dp\,dt}=\frac{g_i^N}{2\pi^2}\frac{\sigma^{1/2}_{\rm abs}(M, M_{N_i}, p)\,}{\exp[E_i(p)/T_{\rm BH}]+1}\frac{p^3}{E_i(p)}\,,
\end{align}
where $g_i^N$ are the internal degrees-of-freedom (dof) of right-handed neutrinos, $\sigma^{1/2}_{\rm abs}(M, M_{N_i}, p)$ is the absorption cross section for a massive fermion  \cite{Page:1976df,Unruh:1976fm,Doran:2005vm}, $E_i(p)$ the right-handed neutrino total energy, and $T_{\rm BH}$ the instantaneous BH temperature. In the case of a Schwarzschild BH, its temperature $T_{\rm BH}$ is related to its mass $M$ as
\begin{align} \label{eq:TBH}
    T_{\rm BH}=\frac{1}{8\pi G M}\approx 1.06\ {\rm GeV} \left(\frac{10^{13}\ {\rm g}}{M}\right)\,.
\end{align}
For BH temperatures $T_{\rm BH} \ll M_{N_i}$, we observe that the emission of right-handed neutrinos becomes highly suppressed. This suppression will impose a constraint on the range of BH masses that will contribute in a significant manner to leptogenesis. 

As a result of the evaporation process, BHs lose mass over time at a rate~\cite{MacGibbon:1990zk,MacGibbon:1991tj},
\begin{align}\label{eq:MEq}
	\frac{dM}{dt}&=-\sum_{a}\frac{g_a}{2\pi^2}\int_0^\infty \frac{\sigma^{s_a}_{\rm abs}(G M p)\,p^3\,dp}{\exp[E_a(p)/T_{\rm BH}]-(-1)^{2s_a}}\,,\notag\\
	&=-\kappa\,\varepsilon(M)\left(\frac{\rm 1\ g}{M}\right)^{2}\,,
\end{align}
where the sum is performed over all existing particles with spin $s_a$, dofs $g_a$ and absorption cross section $\sigma^{s_a}_{\rm abs}(G M p)$. The mass lose rate in Eq.\ \eqref{eq:MEq} is conveniently parametrized in terms of an \textit{evaporation} function, $\varepsilon(M)$, normalized to the unity for $M_{\rm}\gg 10^{17}\g$ where $\kappa = 5.34\times 10^{25} ~{\rm g\, s^{-1}}$ \cite{MacGibbon:1990zk,MacGibbon:1991tj}. For the case of type-I seesaw framework, the evaporation function can be separated into SM plus right-handed neutrino contributions, $\varepsilon(M)=\varepsilon_{\rm SM}(M)+\varepsilon_{N}(M)$~\cite{Lunardini:2019zob}:
 \begin{align}
    \varepsilon_{N}(M) \approx 2\,n_{N_{i}} f_{1/2}^{ 0}\sum_{i=1}^{n_{N_{i}}}\exp\left[-\frac{8\pi GMM_{N_i}}{4.53} \right]\,,
\end{align}
where  $n_{N_{i}}$ denotes the number of generations of  right-handed neutrinos, which we assume to be three and the factor $f_{1/2}^{ 0}$ corresponds to the neutral fermion contribution to the evaporation function~\cite{MacGibbon:1991tj}. The presence of the exponential factor characterizes the BH mass where the right-handed neutrino emission becomes suppressed. 

%%%%%%%%%%%%%%%%%%%%%%%%%%%%%%%%%%%%%%%%%%%%%%%%%%%%%%%%%%%
%%%%%%                                    Boltzmann equations                                 %%%%%
%%%%%%%%%%%%%%%%%%%%%%%%%%%%%%%%%%%%%%%%%%%%%%%%%%%%%%%%%%%
\textit{The interplay between leptogenesis and early black hole domination.} --- 
Let us assume an initial density of PBHs, $\rho_{\rm PBH}^i$, formed after inflation with a monochromatic mass distribution. The initial PBH mass is proportional to the initial radiation density at the formation time, $M_i \propto \gamma \rho_{\rm tot}^i H_i^{-3}$, where $\gamma$ is a dimensionless gravitational collapse parameter and $H_i$ the Hubble parameter~\cite{Carr:2020gox}. The ratio of the initial PBH fraction to the total energy, $\rho^i_{\rm tot}$, for a given plasma temperature at the formation $T_{\rm f}$, is~\cite{Carr:2020gox}
\begin{align}
	\beta^\prime=\gamma^{1/2}\left(\frac{g_*(T_{\rm f})}{106.75}\right)^{-1/4}\frac{\rho_{\rm PBH}^i}{\rho_{\rm tot}^i}\,.
\end{align} 
Depending on the value of $\beta^\prime$, the PBHs could eventually dominate the evolution of the universe before their evaporation. 
We consider PBHs in the mass range ($10^{-1}\lesssim M_{i} \left(\text{g}\right)\lesssim 10^4$) where the upper boundary derives from PBH evaporation that would occur before the electroweak phase transition and the lower from constraints on inflation \cite{Masina:2020xhk,Baldes:2020nuv}.

If the type-I seesaw mechanism explains the tininess of neutrino masses, then right-handed neutrinos would be produced in the early universe. 
The right-handed neutrinos would be produced by the BHs when their temperature $T_{\rm BH} \ll M_{N_1}$  and also from the thermal plasma itself.
To calculate the resultant baryon asymmetry we track the evolution of the comoving number density of the right-handed neutrinos, primordial black holes, lepton asymmetry and the radiation energy density using momentum integrated Boltzmann equations. These equations assume that kinetic equilibrium of the right-handed neutrinos is achieved and neglects quantum statistics. We find that for the points in the model parameter space
we explore,  leptogenesis occurs in the strong washout regime where it has been shown that momentum averaged Boltzmann equations yields a near-identical lepton asymmetry compared to the full treatment \cite{HahnWoernle:2009qn}.

Further, we consider the viability of non-resonant thermal leptogenesis and assume a mass splitting of $M_{2}=3.15\, M_{1}$ and $M_{3}=3.15\,M_{2}$ and  $10^{6}\lesssim M_1 \left(\text{GeV}\right)\lesssim 10^{13}$. The upper and lower boundary reflects the viable region for non-resonant thermal.
 We begin with the following Friedmann equations for the comoving energy density of radiation ($\varrho_{\rm R}= a^4 \rho_{\rm R}$) and PBHs ($\varrho_{\rm BH}= a^3 \rho_{\rm BH}$) with respect to the scale factor $a$:
\begin{subequations}\label{eq:UnEv}
\begin{align}
  	\frac{d\varrho_{\rm R}}{da} &= \frac{1}{a\Delta}\left[4(\Delta-1)-\Sigma\right]-\frac{\varepsilon_{\rm SM}(M)}{\varepsilon(M)}\frac{1}{M}\frac{dM}{da}a\varrho_{\rm BH}\,,\label{eq:UnEvRad}\\ 
    \frac{d\varrho_{\rm BH}}{da} &=\frac{1}{M}\frac{dM}{da} \varrho_{\rm BH}\,,\\
    H^2&=\frac{8\pi G}{3}\left(\varrho_{\rm BH} a^{-3}+\varrho_{\rm R} a^{-4}\right)\,,
\end{align}
\end{subequations}
where $H$ is the Hubble rate. The first term in Eq.~\eqref{eq:UnEvRad} takes into account the change on the effective number of relativistic degrees of freedom $g_*(T)$ as the Universe cools down, where $\Sigma$ and $\Delta$ functions are defined as
\begin{align}
    \Sigma = \frac{T}{g_*(T)}\frac{dg_*(T)}{dT},\quad \Delta = 1 + \frac{T}{3 g_{*S}(T)}\frac{dg_{*S}(T)}{dT}.
\end{align}
As a result of the PBH evaporation, the entropy of the universe is not conserved. Therefore, we need to follow the evolution of the ambient temperature, $T$, using~\cite{Lunardini:2019zob, Bernal:2019lpc,Arias:2019uol}
\begin{align}\label{eq:TUev}
\frac{dT}{da} &= -\frac{T}{\Delta}\left\{ \frac{1}{a} + \frac{\varepsilon_{\rm SM}(M)}{\varepsilon (M)}\frac{1}{M}\frac{dM}{da} \frac{g_{*}(T)}{g_{*S}(T)} \frac{a \varrho_{\rm BH}}{4\varrho_{\rm R}}\right\}~.
\end{align}
We emphasize that the ambient plasma temperature, $T$, is not necessarily the same as the temperature of the black holes, $T_{\rm BH}$.
To address the generation of baryon asymmetry, we require Boltzmann equations for the evolution of the comoving right-handed neutrino number density and the $B-L$ asymmetry produced from the decays (and washout) of the right-handed neutrinos. 
We assume a vanishing initial abundance of right-handed neutrinos. They are populated 
thermally, from inverse decays of leptons and Higgses, and non-thermally from the Hawking radiation. 
For sake of clarity, let us separate the equations for the thermal ($n_{N_1}^{\rm TH}$) and non-thermal ($n_{N_1}^{\rm BH}$) densities\footnote{We restrain from evolving with respect to the usual $z=M_{N_1}/T$ parameter since the entropy is not conserved \cite{Buchmuller:2011mw}.},
\begin{subequations}\label{eq:BEright-handedn}
\begin{align}
	a H \frac{dn_{N_1}^{\rm TH}}{da} &= -(n_{N_1}^{\rm TH}-n_{N_1}^{\rm eq})\Gamma_{N_1}^T\,,\label{eq:BERH-TH}\\
	a H \frac{dn_{N_1}^{\rm BH}}{da} &= -n_{N_1}^{\rm BH}\Gamma_{N_1}^{\rm BH}+ n_{\rm BH} \Gamma_{\rm BH\to N_1}\,,\label{eq:BERH-BH}
\end{align}
\end{subequations}
where $\Gamma_{N_1}^T$ and $n_{N_1}^{\rm eq}$ are the thermally averaged decay rate and the equilibrium abundance of the right-handed neutrinos, respectively. $\Gamma_{N_1}^{\rm BH}$ in Eq.~\eqref{eq:BERH-BH} is the decay width corrected by an average inverse time dilatation factor
\begin{align}\label{eq:GBH}
	\Gamma_{N_1}^{\rm BH} \equiv \left\langle\frac{M_{N_1}}{E_{N_1}}\right\rangle_{\rm BH} \Gamma_{N_1}^{0} \approx \frac{{\cal K}_1(z_{\rm BH})}{{\cal K}_2(z_{\rm BH})} \Gamma_{N_1}^0\,,
\end{align}
where $\Gamma_{N_1}^0$ is the right-handed neutrino decay width, ${\cal K}_{1,2}(z)$ are modified Bessel functions of the second kind, and we defined $z_{\rm BH}=M_{N_1}/T_{\rm BH}$. The average is taken with respect to the BH instantaneous spectrum since the right-handed neutrino energies are distributed according to the Hawking rate, which resembles a thermal distribution. The approximated form of $\Gamma_{N_1}^{\rm BH}$ as a function of ${\cal K}_{1,2}(z)$ is obtained assuming that the Hawking spectrum has a Maxwell-Boltzmann form. 
 
In our numerical code we use the full greybody factors, nonetheless. Eq.~\eqref{eq:BERH-BH} contains a source term related to the PBH evaporation, equal to the PBH number density, $n_{\rm BH}\equiv \varrho_{\rm BH}/M$, times $\Gamma_{\rm BH\to N_1}$, the right-handed neutrino emission rate per BH. Integrating the Hawking rate, Eq.~\eqref{eq:nuHS}, with respect to the momentum, we obtain such a rate~\cite{Page:1977um,Page:2007yr,MacGibbon:2007yq},
\begin{align}\label{eq:BHdec}
\Gamma_{\rm BH\to N_1}&\equiv \int_0^\infty \frac{d^2 {\cal N}_{N_1}}{dp\,dt} dp\,,\notag\\
& \approx \frac{27 T_{\rm BH}}{32\pi^2}\left(-z_{\rm BH} {\rm Li}_2(-e^{-z_{\rm BH}}) - {\rm Li}_3(-e^{-z_{\rm BH}})\right)\,,
\end{align}
where ${\rm Li}_s(z)$ are polylogarithm functions of order $s$; assuming the fermion greybody factor equal to the geometric optics limit,  $\sigma^{1/2}_{\rm abs}(M, M_{N_i}, p) = 27 \pi G^2M^2$, we obtain such analytical approximation.
\begin{figure*}[t!]
\includegraphics[width=0.95\textwidth]{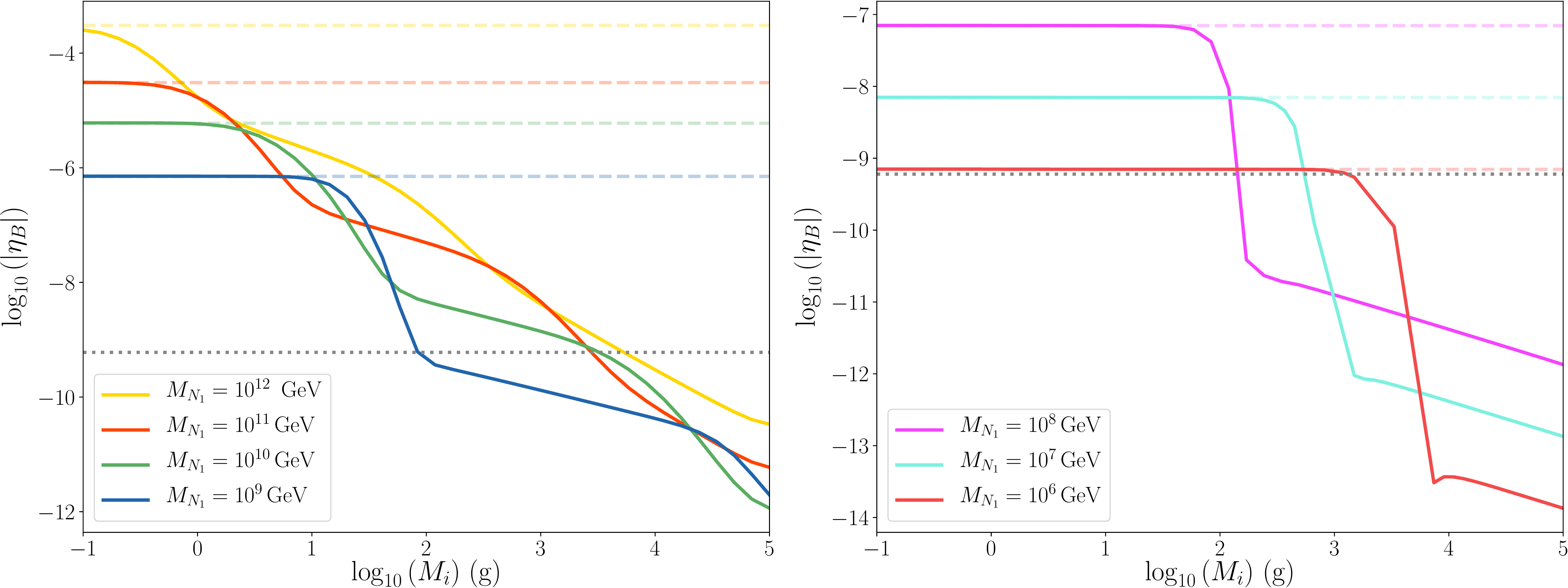}
\caption{The final baryon-to-photon ratio as a function of the initial black hole masses for varying masses of the right-handed neutrino. The grey dotted line shows the measured baryon-to-photon ratio measured from cosmic microwave background radiation data \cite{Ade:2015xua} while the color dashed lines indicate the result obtained from leptogenesis in the case that there was no PBH dominated era. In both plots we assume $\beta^{\prime}=10^{-3}$.\label{fig:etaBMi}}
\end{figure*}

The equation for the $B-L$ asymmetry, $n^{\rm B-L}_{\alpha\beta}$, reads
\begin{align}\label{eq:BElep}
	a H \frac{dn^{\rm B-L}_{\alpha\beta} }{da}&= \epsilon_{\alpha\beta}^{(1)}\left[(n_{N_1}^{\rm TH}-n_{N_1}^{\rm eq})\Gamma_{N_1}^T + n_{N_1}^{\rm BH} \Gamma_{N_1}^{\rm BH}\right]+ {\cal W}_{\alpha\beta}\,,
\end{align}
where $\epsilon_{\alpha\beta}^{(1)}$ is the CP-asymmetry matrix describing the decay asymmetry generated by  $N_{1}$, $ {\cal W}_{\alpha\beta}$ are the washout terms, which we detail in the  Supplemental Material. The generation of the lepton asymmetry has thermal and non-thermal sources stemming from the plasma and PBH evaporation respectively. However, the washout is independent of the PBHs. 

Similar to the thermal leptogenesis scenario, the PBH induced leptogenesis can be understood through the $z_{\rm BH}$ parameter, cf.~Eqs.~\eqref{eq:GBH} and \eqref{eq:BHdec}. In the case that $z_{\rm BH}\gg 1$, the right-handed neutrino emission becomes suppressed as the right-handed neutrino mass far exceeds the temperature of the black hole, and the effect on the final asymmetry is negligible. On the other hand, if $z_{\rm BH} \lesssim 1$, the PBHs emit right-handed neutrinos independent of the conditions of the surrounding thermal plasma.  However, the final contribution to the baryon asymmetry is crucially dependent on the CP-asymmetry matrix ($\epsilon_{\alpha\beta}$), related to the Yukawa couplings, and on the properties of the ambient plasma. Thus, according to the period when the black hole induced leptogenesis is active relative to the thermal leptogenesis era ($z = M_{N}/T\sim \mathcal{O}(1)$), the PBH contribution can be substantial or not. For instance, if washout processes are ineffective when the right-handed neutrinos are emitted by the PBHs ($z \gtrsim 10$), their contribution to the final baryon-to-photon ratio can be enhanced. 
Nevertheless, there is another additional effect acting during the baryon asymmetry generation: PBHs also produce a significant population of photons, effectively reducing the baryon-to-photon ratio.  This interplay will ultimately determine the final baryon asymmetry.

%%%%%%%%%%%%%%%%%%%%%%%%%%%%%%%%%%%%%%%%%%%%%%%%%%%%%%%%%%%%%%%%%%%%%%%%%%%
%\section{RESULTS AND DISCUSSION}
%%%%%%%%%%%%%%%%%%%%%%%%%%%%%%%%%%%%%%%%%%%%%%%%%%%%%%%%%%%%%%%%%%%%%%%%%%%
\textit{Results and discussion.} --- 
We solve the system of equations \eqref{eq:UnEv}, \eqref{eq:TUev}, \eqref{eq:BEright-handedn}, and \eqref{eq:BElep}, together with the PBH mass rate, Eq.\ \eqref{eq:MEq}, to obtain the final baryon asymmetry in a PBH dominated universe. For the sake of generality, we do not include any bound on the initial PBH fraction.  We have implemented such system of equations in a plugin to be used with the {\tt ULYSSES} python package \cite{Granelli:2020pim}. In \figref{fig:etaBMi} we present the baryon-to-photon ratio $\lvert \eta_{\rm B}\rvert$ as a function of the initial PBH mass $M_i$, for some specific sets of right-handed neutrino masses and Yukawa matrix elements. We assume $\beta^\prime=10^{-3}$ such that PBH domination is guaranteed. We will discuss the dependence on the initial fraction later on. 
Furthermore, the parameters related to the Yukawa matrix are provided in the Supplemental Material and are chosen such that the baryon asymmetry will be maximally enhanced while preserving the perturbative series of the neutrino masses \cite{Moffat:2018wke}. Therefore, the dilutionary effect we observe would only be more pronounced in other parameter space regions.
\begin{figure*}[t!]
\includegraphics[width=\textwidth]{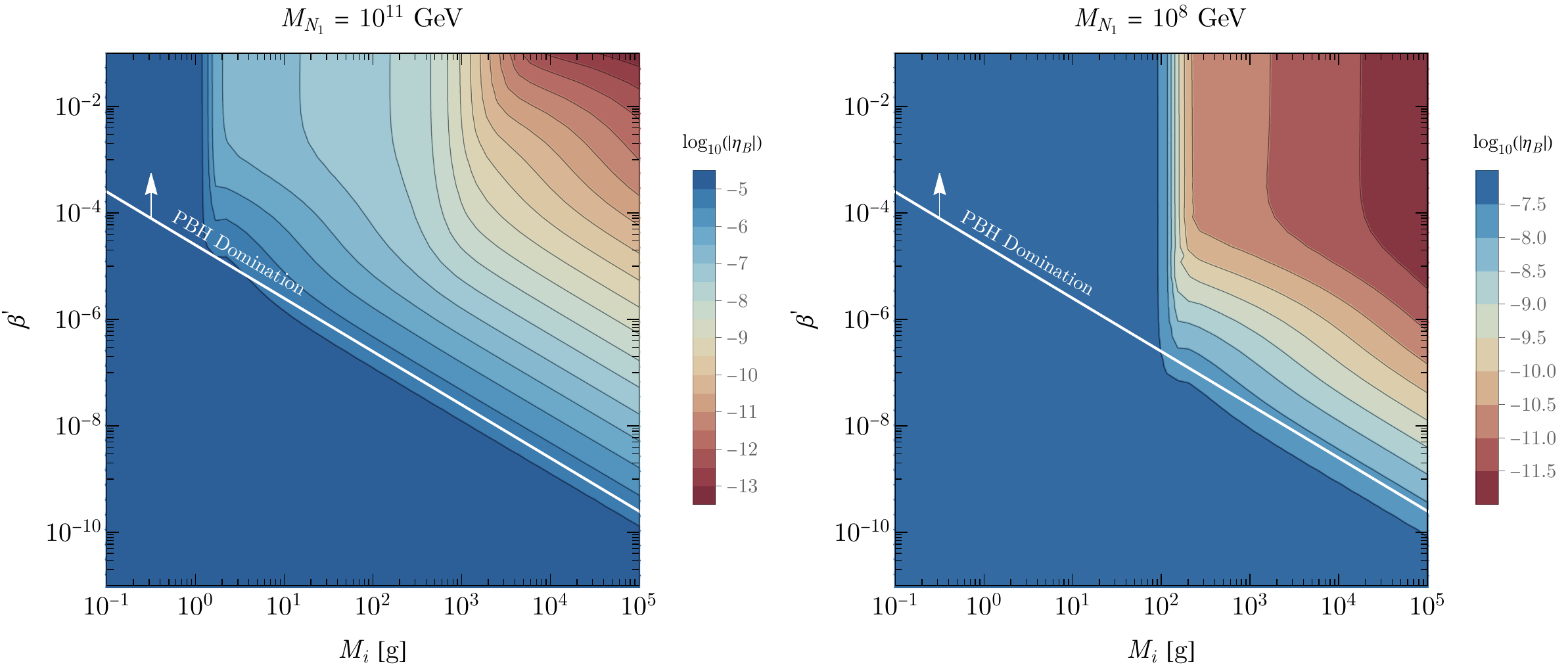}
\caption{$|\eta_{\rm B}|$ as function of the initial fraction $\beta^\prime$ and PBH mass $M_i$ for two different benchmarks, $M_{N_1}=10^{11}\GeV$ (left) and $M_{N_1}=10^{8}\GeV$ (right). The white line indicates the values from which it is expected an early PBH dominated era.}\label{fig:bpvsMi}
\end{figure*}

In the case that the thermal leptogenesis era occurs before the PBH evaporation, we observe that the entropy injection largely dilutes the initial baryon-to-photon ratio. Nevertheless, depending on the right-handed neutrinos and initial PBH masses' specific values, the PBH induced asymmetry can partially mitigate the depletion of the final baryon-to-photon ratio. In the high mass regime ($10^9\GeV \lesssim M_{N_1} \lesssim 10^{12}\GeV$, Fig~\ref{fig:etaBMi}, left panel), we observe the depletion of baryon asymmetry and the limited mitigation from the PBH-induced leptogenesis depending on the initial PBH mass. Let us consider as benchmark the case of $M_{N_1}=10^{11}\GeV$ (red line). For $M_i\lesssim 1\g$, the final baryon-to-photon ratio is only marginally affected because the evaporation takes places before the thermal leptogenesis era. Thus, the right-handed neutrinos produced by the PBHs only constitute an initial condition for the leptogenesis. For $M_i\gtrsim 1\g$, the PBHs start to inject a large quantity of radiation, reducing significantly $\lvert \eta_{\rm B}\rvert$.  However, for $1\g\lesssim M_i \lesssim 3\times 10^3\g$, the contribution coming from the right-handed neutrinos emitted by the PBHs ameliorate the diminution of the final baryon asymmetry. As such additional right-handed neutrinos are emitted after thermal leptogenesis, they do not experience strong washout effects. Therefore, as noted before, their out-of-equilibrium decays alleviate the reduction of the baryon asymmetry. This effect is maximal when $z_{\rm BH} \sim 1$, \ie, 
\begin{align}\label{eq:MMi}
	M_i \sim \frac{1}{8\pi G M_{N_1}} \sim 105.7\g\left(\frac{10^{11}\GeV}{M_{N_1}}\right)\,.
\end{align}
For PBH masses larger than such a value, the right-handed emission becomes suppressed, so that the only effect present corresponds to the reheating of the primordial plasma. In fact, for $M_i\sim 10^4\g$, $\lvert \eta_{\rm B}\rvert$ is decreased by a factor of $\sim{\cal O}(10^6)$ with respect to minimal thermal leptogenesis scenario. For other values of the right-handed neutrino masses, similar behavior of  $\lvert \eta_{\rm B}\rvert$ is observed. The partial mitigation due to the PBH-induced asymmetry is shifted in $M_i$ according to Eq.~\eqref{eq:MMi} and becomes less pronounced because the Yukawa couplings are smaller for lower right-handed neutrino masses, cf.~Eq.~\eqref{eq:Yuk}.
For the intermediate-mass regime, $10^6\GeV \lesssim M_{N_1} \lesssim 10^{9}\GeV$ (Fig~\ref{fig:etaBMi}, right panel), the injected entropy largely affects the total $\lvert \eta_{\rm B}\rvert$ for initial PBH masses $M_i\gtrsim 10^2\g\ (3\times 10^3\g)$ in the case of right-handed neutrino masses $M_{N_1}=10^8\GeV\ (10^6\GeV)$. The mitigation present in the high-mass regime is absent here, as the contribution from PBH evaporation to the total $\lvert \eta_{\rm B}\rvert$ is reduced due to their smaller Yukawa couplings.

In Fig.~\ref{fig:bpvsMi}, we present the dependence of the final baryon-to-photon ratio on the initial PBH fraction $\beta^\prime$ and mass $M_i$ for two benchmark cases of the right-handed neutrino parameters -- $M_{N_1}=\{10^8,10^{11}\}\GeV$. In the high mass regime, PBHs with masses $M_i\lesssim 1\g$ do not affect the final baryon asymmetry, independently of $\beta^\prime$, as noted previously. For $M_i\gtrsim 1\g$, we can distinguish two generic regions.  If $\beta^\prime \gtrsim 10^{-4}$, we find a similar behavior as observed in Fig.~\ref{fig:etaBMi}: an increasing depletion of $|\eta_{\rm B}|$ for larger PBH masses, partially mitigated only for values in which $z_{\rm BH}\lesssim 1$. For lower values of $\beta^\prime$, the reduction caused by the entropy injection is more dependent on the initial fraction. However, let us notice that, within the region leading to a PBH dominated era, the asymmetry is at least reduced by a factor of $\sim 5$. Similar features are present in the intermediate-mass regime benchmark. For $\beta^\prime \gtrsim 10^{-4}$, the depletion decreases the asymmetry at least by a factor of $\sim{\cal O}(10^3)$ for $M_i\gtrsim 10^2\g$, independently of $\beta^\prime$. For smaller initial PBH fractions, the entropy injection diminishes the baryon-to-photon ratio by a factor between $\sim 5 - 10^3$ depending on the specific PBH initial fraction and mass. Nonetheless, let us stress that even the reduction by a factor of 5 induces a tension on the intermediate-mass regime.

%%%%%%%%%%%%%%%%%%%%%%%%%%%%%%%%%%%%%%%%%%%%%%%%%%%%%%%%%%
%%%%%%                        GRAVITATIONAL WAVES                             						       %%%%%%
%%%%%%%%%%%%%%%%%%%%%%%%%%%%%%%%%%%%%%%%%%%%%%%%%%%%%%%%%%

\textit{Gravitational wave signature.} --- 
From the recent advances in multi-messenger observations, we may wonder whether it is possible to demonstrate that existed an early PBH dominated era. Diverse studies have shown the possibility of testing leptogenesis and/or PBHs considering gravitational waves (GW) \cite{Dolgov:2011cq,Dror:2019syi,Blasi:2020wpy,Samanta:2020cdk,Inomata:2020lmk,Ejlli:2019bqj}. Such GWs can be produced from different phenomena~\cite{Hooper:2020evu,Dolgov:2011cq,Dong:2015yjs}. The GW from direct Hawking emission, for instance, have significantly high frequencies ---${\cal O}(f_0)\sim 100$ THz for $M_i=10^{-1}\g$--- out of reach for current detectors (see the Supplemental Material for further details). Nonetheless, there are numerous experimental proposals to probe high-frequency regimes \cite{Arvanitaki:2012cn,Ito:2019wcb,Chen:2020ler,Chou:2015sle,Page:2020zbr}. Thus, we conclude that if evidence of a PBH dominated early universe consistent with initial PBH masses of ${\cal O}(M_i)\gtrsim 1{\rm\ kg}$ are found, this would place significant tension on the intermediate-scale leptogenesis scenario.

%%%%%%%%%%%%%%%%%%%%%%%%%%%%%%%%%%%%%%%%%%%%%%%%%%%%%%%%%%%%%%%%
%%%%%%                                   Acknowledgments										          %%%%%%
%%%%%%%%%%%%%%%%%%%%%%%%%%%%%%%%%%%%%%%%%%%%%%%%%%%%%%%%%%%%%%%%
\acknowledgments

\textit{Acknowledgments.} --- 
We want to thank Cecilia Lunardini for her involvement during the initial stages of this project. 
We are grateful to Kristian Moffat for useful discussions on leptogenesis. Fermilab is operated by the Fermi Research Alliance, LLC under contract No. DE-AC02-07CH11359 with the United States Department of Energy.

\textit{Note added:} During the end stage preparation of this work, a paper by D. Hooper and G. Krnjaic appeared where they also consider primordial black hole induced leptogenesis~\cite{Hooper:2020otu}. 

\bibliographystyle{apsrev4-1}
\bibliography{ref}{}

\clearpage
\onecolumngrid
\appendix

\section{Supplemental Material}

In this Supplemental Material, we provide technical details, which may be relevant to experts, some additional examples with figures of the PBH induced leptogenesis and give explicit values of the parameters used in the figures of the main text.

\section{A. Friedmann - Boltzmann Equations}

Since Friedmann equations for the evolution of the Universe cover a huge range in the scale factor $a$, in our code we evolve the system of equations with respect to the logarithm of $a$, $\alpha \equiv \log(a)$. Here, we provide explicitly the system of equations (5), (8), and (9) as function of $\alpha$
\begin{subequations}\label{eq:BEnolog}
\begin{align}
    \frac{dM}{d\alpha}&= -\frac{\kappa}{H}\,\varepsilon(M)\left(\frac{\rm 1\ g}{M}\right)^{2} \,,\\
    \frac{d\varrho_{\rm R}}{d\alpha} &= -\frac{\varepsilon_{\rm SM}(M)}{\varepsilon(M)}\frac{1}{M}\frac{dM}{d\alpha}\varrho_{\rm BH}\,,\\
    \frac{d\varrho_{\rm BH}}{d\alpha} &= \frac{1}{M}\frac{dM}{d\alpha}10^\alpha\varrho_{\rm BH}\,,\\
    \frac{dT}{d\alpha} &= -\frac{T}{\Delta}\left\{ 1 + \frac{\varepsilon_{\rm SM}(M)}{\varepsilon_{\rm D}(M)}\frac{1}{M} \frac{dM}{d\alpha} \frac{g_{*}(T)}{g_{*S}(T)} \frac{10^\alpha \varrho_{\rm PBH}}{4\varrho_{N}}\right\}\,,\\
    H^2&=\frac{8\pi G}{3}\left(\varrho_{\rm BH} 10^{-3\alpha}+\varrho_{\rm R} 10^{-4\alpha}\right)\,,
\end{align}
\end{subequations}

The Boltzmann equations for the evolution of the RH neutrino number density and the lepton asymmetry in Eqs.~(11), (14) as function of $\alpha$, explicitly presenting the washout terms for the three flavour components, $e, \mu, \tau$, are given by
\begin{subequations}\label{eq:BELeplog}
\begin{align}
\frac{dn_{N_1}^{\rm TH}}{d\alpha} &= -(n_{N_1}^{\rm TH}-n_{N_1}^{\rm eq})\frac{\Gamma_{N_1}^T}{H}\,,\\
\frac{dn_{N_1}^{\rm BH}}{d\alpha} &= -n_{N_1}^{\rm BH}\frac{\Gamma_{N_1}^{\rm BH}}{H}+ \frac{\Gamma_{\rm BH\to N_1}}{H}\frac{\varrho_{\rm BH}}{M}\,,\\
   \frac{d N^{\rm B-L}_{\alpha\beta}}{d\alpha} &= \epsilon_{\alpha\beta}^{(1)}\left[(n_{N_1}^{\rm TH}-n_{N_1}^{\rm eq})\frac{\Gamma_{N_1}^T}{H} + n_{N_1}^{\rm BH}\frac{\Gamma_{N_1}^{\rm BH}}{H}\right] -\frac{1}{2}\frac{W_1}{H}\left\{P^{0(1)},N^{\rm B-L}\right\}_{\alpha\beta}\notag\\
    &\quad -\frac{\Gamma_\mu}{2H}\left[
    \begin{pmatrix}
    1 & 0 & 0\\
    0 & 0 & 0\\
    0 & 0 & 0\\
    \end{pmatrix},
    \left[\begin{pmatrix}
    1 & 0 & 0\\
    0 & 0 & 0\\
    0 & 0 & 0\\
    \end{pmatrix},N^{\rm B-L}\right]
    \right]_{\alpha\beta}
    -\frac{\Gamma_\tau}{2H}\left[
    \begin{pmatrix}
    0 & 0 & 0\\
    0 & 1 & 0\\
    0 & 0 & 0\\
    \end{pmatrix},
    \left[\begin{pmatrix}
    0 & 0 & 0\\
    0 & 1 & 0\\
    0 & 0 & 0\\
    \end{pmatrix},N^{\rm B-L}\right]
    \right]_{\alpha\beta}\,,
\end{align}
\end{subequations}
where $W_1$ is the washout related to the $N_1$ neutrino, 
\begin{align}
	W_1=\frac{1}{4}\Gamma_{N_1}^T{\cal K}_2(z)z^2\,,
\end{align}
and $P^{0(i)}\equiv c_{i\alpha}c_{i\beta}^*$ are projection matrices describing the washout of a specific flavor \cite{Blanchet:2011xq}. 
We solve this system of equations using {\tt ULYSSES} python package \cite{Granelli:2020pim}. However, we do not
use the default conversion for the baryon-to-photon ratio (which assumes entropy conservation). Rather we normalize by the comoving
number density of photons as outlined in our plugin. For a detailed discussion of  Boltzmann equations which incorporates
entropy production see Ref.~\cite{Buchmuller:2011mw}.

\section{B. Prototypical solutions to PBH-induced leptogenesis}

Understanding the behaviour of the PBH-induced leptogenesis involves solving the system of coupled equations presented in \eqref{eq:BEnolog} and \eqref{eq:BELeplog}. 
The enhancement and depletion of the baryon asymmetry, for differing PBHs and right-handed neutrino masses, cannot be captured with simple, closed analytic expressions. Here we
show specifically the results for three benchmark points, labeled ``before'', ``during'' and ``after'' according to when the PBH disappearance occurs relative to thermal 
leptogenesis. We consider a region of the parameter space where the lepton asymmetry is enhanced due to fine-tuning of the R-matrix while  retaining the
perturbativity of the Yukawa couplings. The R-matrix has the 
following form,
\begin{equation}
R=\begin{pmatrix}
1 & 0 & 0 \\
0 & c_{\omega_{1}} & s_{\omega_{1}} \\
0 &- s_{\omega_{1}} & c_{\omega_{1}} 
\end{pmatrix}
\begin{pmatrix}
c_{\omega_{2}} & 0 & s_{\omega_{2}} \\
0 & 1 & 0\\
-s_{\omega_{2}} & 0 & c_{\omega_{2}} 
\end{pmatrix}\\
\begin{pmatrix}
c_{\omega_{3}} & s_{\omega_{3}} & 0\\
-s_{\omega_{3}} & c_{\omega_{3}} & 0\\
0 & 0 & 1
\end{pmatrix},
\end{equation}
where $c_{\omega_{i}} \equiv \cos\omega_{i}$, $s_{\omega_{i}} \equiv \sin\omega_{i}$  and the complex angles are given by $\omega_{i} \equiv x_{i}+iy_{i}$ with $\lvert x_{i}\rvert, \lvert y_{i}\rvert\leq180^{\circ}$ for $i= 1, 2, 3$.

The parameters of the Yukawa matrix are determined by the Casas-Ibarra parametrisation. We choose for this benchmark point the following values,
\begin{align}\label{eq:ParY}
 m_{1}		&=0.12\,\rm{eV},          &  M_{N_{1}}&=10^{11}\,\rm{GeV},      &  M_{N_{2}}&=3.16\times 10^{11}\,\rm{GeV},        & M_{N_{3}}  &=10^{12}\,\rm{GeV},\notag\\
x_{1} 		& = 132.2^{\circ},          &    y_{1}      &= 175.0^{\circ},             & x_{2}         & = 87.8^{\circ},                                  &          y_{2} &= 2.9^{\circ}, \\
x_{3}      	& = -30.3^{\circ},          &    y_{3}       &= 175.0^{\circ},             &  \delta       &=281.2^{\circ},                                 & \alpha_{21} & = 181.90^{\circ},  \notag\\
\alpha_{31} & = 344.7^{\circ},          &   \theta_{23}&= 46.2^{\circ},              &  \theta_{12} &=33.8^{\circ},                                 & \theta_{13} & = 8.61^{\circ},  \notag\\
\end{align}
where we assume normal ordering and the masses of $m_{2}$ and $m_{3}$ are fixed from the best fit values of the mass squared splittings taken from ~\cite{Esteban:2020cvm}.
The ``before'', ``during'' and ``after'' PBH masses are $10^{-1}$, $1$  and $10^4$ g respectively. For each scenario, we take as the initial PBH fraction $\beta^\prime=10^{-3}$, such that it gives a PBH domination at some point in the evolution of the Universe. 

\begin{figure}[t!]
\includegraphics[width=\textwidth]{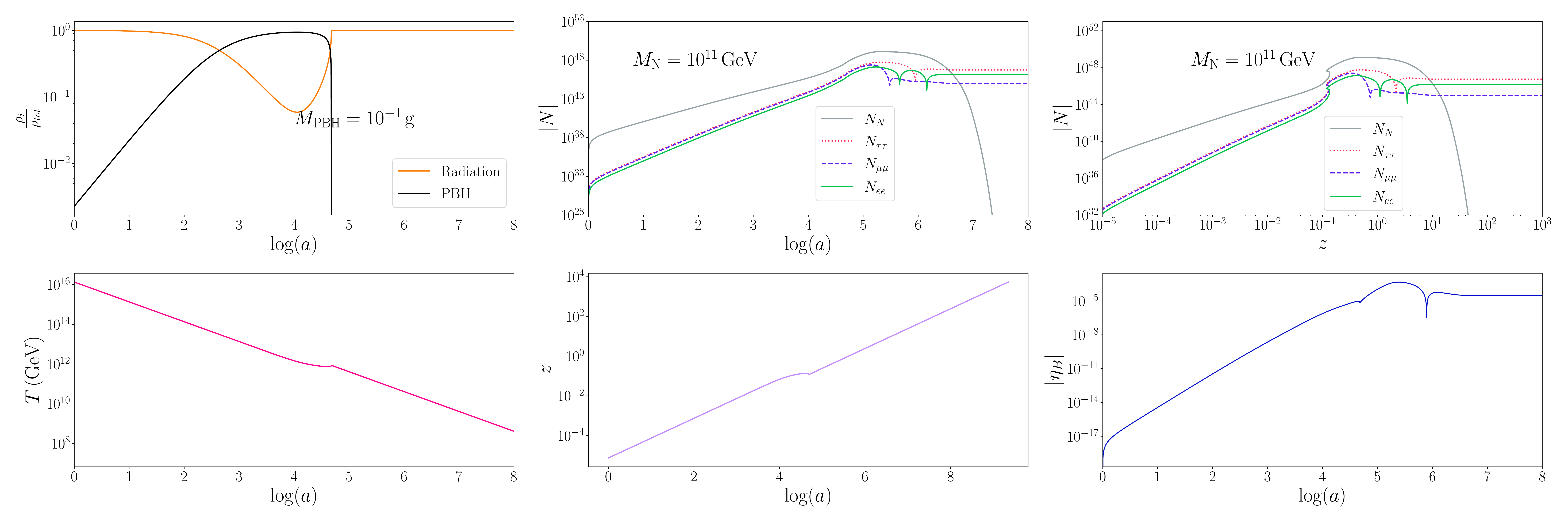}\\
\caption{The upper left plot shows the evolution of energy densities of the black holes (black) and radiation (yellow) as a function of the logarithm of the
scale factor for the ``before'' scenario. The upper central plot shows the evolution of the magnitude of right-handed neutrino and flavoured components of the lepton asymmetry comoving number densities as a function of the logarithm of the scale factor ($a$). The upper right shows the same evolution but as a function of $z=M_{N_{1}}/T$. The temperature of the plasma as a function of the logarithm of the scale factor is shown in the lower-left panel, and the central lower panel shows the
$z=M_{N_{1}}/T$ as a function of $\log(a)$. The lower right plot shows the magnitude of the baryon-to-photon ratio as a function of the logarithm of the scale factor.}\label{fig:before}
\end{figure}
\begin{figure}[t!]
\includegraphics[width=\textwidth]{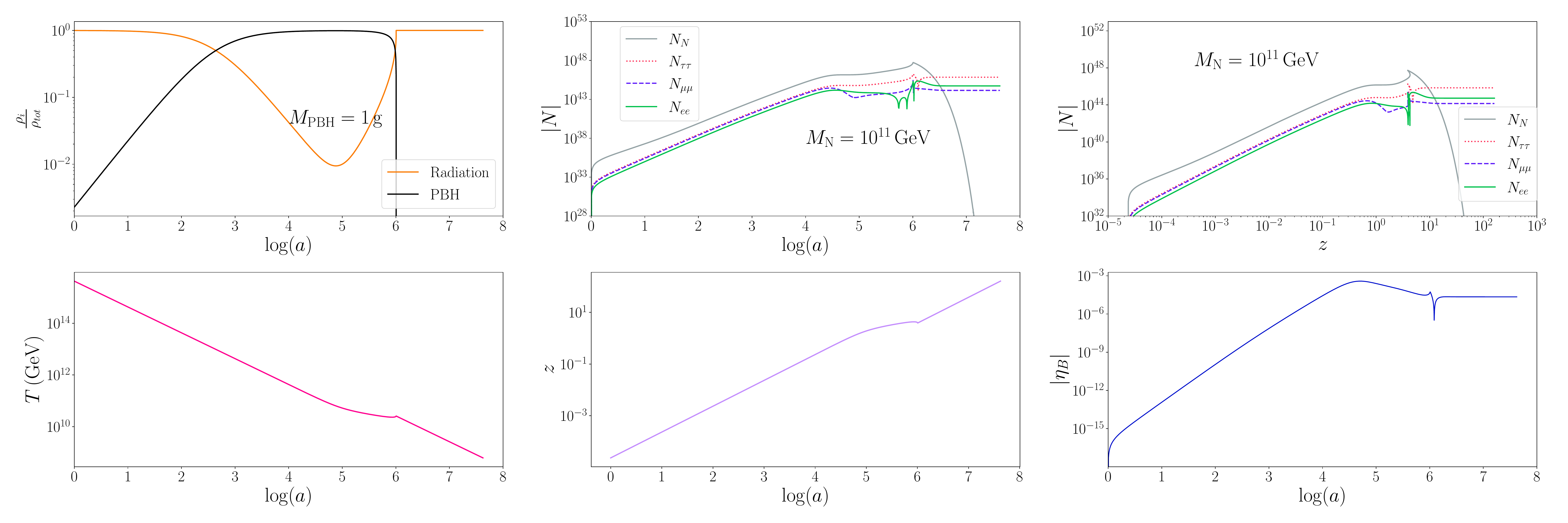}\\
\caption{The analogous plots for the ``during'' scenario.}\label{fig:during}
\end{figure}
\begin{figure}[t!]
\includegraphics[width=\textwidth]{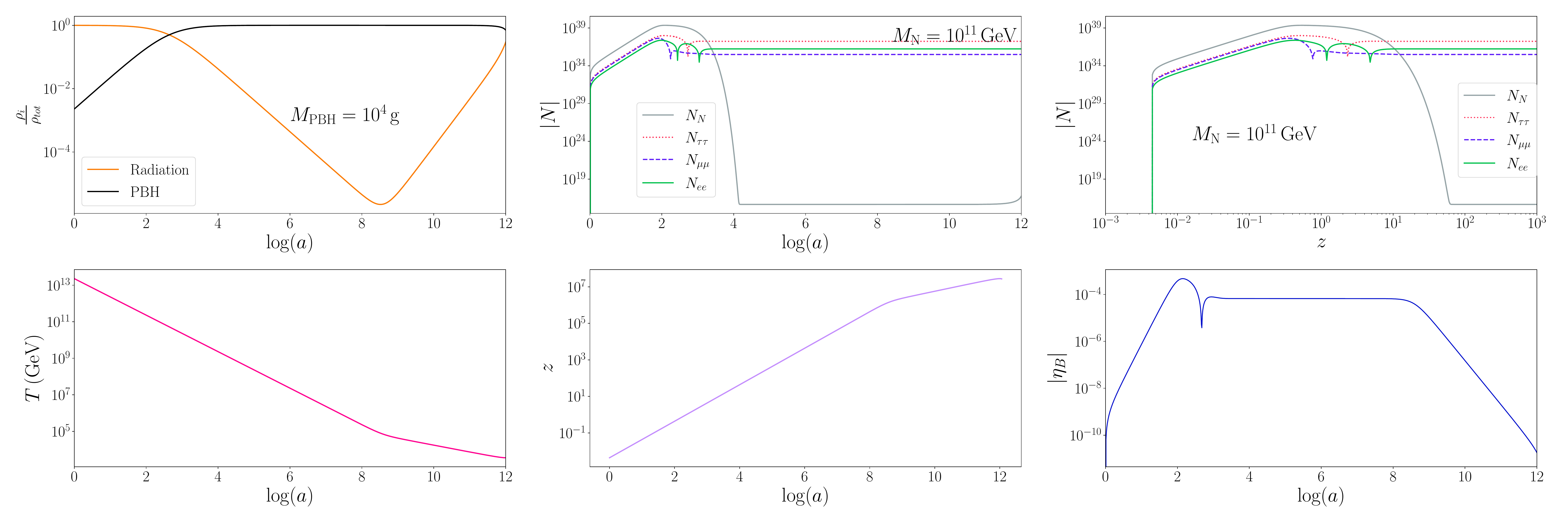}
\caption{The analogous plots for the ``after'' scenario.}\label{fig:after}
\end{figure}
\figref{fig:before} shows the ``before'' scenario where the black hole disappearance occurs around $\log(a)\sim 4.5$. 
We observe the thermal plasma temperature raises slightly at this point (see lower left plot). 
The ``time'' when PBH-induced leptogenesis occurs corresponds to $z\sim 10^{-1}$, and we observe a 
spike in the comoving number density of the right-handed neutrinos in the upper right plot due to the non-thermal contribution from the PBHs.  For $z>1$, PBH induced and thermal leptogenesis has completed, and the lepton asymmetry number density stabilizes. 
\figref{fig:during} shows the ``during'' scenario where the black hole evaporation occurs around $\log(a)\sim 6$ and 
$z\sim 1$. Again, we observe a spike in the comoving number density of the right-handed neutrinos stemming from the explosive evaporation of the PBHs.
This slight increase is not sufficient to overcome the washout processes which are still active in this regime, and the final baryon-to-photon ratio reduces slightly at $\log(a) > 6$.
\figref{fig:after} shows the  ``after''  scenario where the black hole evaporation occurs around $\log(a)\sim 11$ and 
$z\sim 10^{6}$. This is well after the era of thermal leptogenesis. 
We observe that the comoving number density of the right-handed neutrinos falls, due to Boltzmann suppression, until large $\log(a)$ values when the PBHs populate them.
In spite of this, there is a massive entropy dump by the PBH, which produces photons thereby diluting the baryon-to-photon ratio significantly. 

\section{C. RH mass parameters}

Finally, we provide in Tab.~\ref{tab:MassRH} the RH neutrino masses used to obtain the baryon-to-photon ratio as function of the initial PBH mass, Fig. 1. The remaining parameters required by the Casas-Ibarra parametrization to obtain the Yukawa matrizes are the same given in Eq.~\eqref{eq:ParY}.
\begin{table}[h!]
\begin{tabular}{ccc} \toprule
		$M_{N_1}$ & $M_{N_2}$  & $M_{N_3}$ \\  \colrule
		$10^{6}\GeV$ & $3.16\times 10^{6}\GeV$ & $10^{7}\GeV$ \\  \colrule
		$10^{7}\GeV$ & $3.16\times 10^{7}\GeV$ & $10^{8}\GeV$ \\  \colrule
		$10^{8}\GeV$ & $3.16\times 10^{8}\GeV$ & $10^{9}\GeV$ \\  \colrule
		$10^{9}\GeV$ & $3.16\times 10^{9}\GeV$ & $10^{10}\GeV$ \\  \colrule
		$10^{10}\GeV$ & $3.16\times 10^{10}\GeV$ & $10^{11}\GeV$ \\  \colrule
		$10^{11}\GeV$ & $3.16\times 10^{11}\GeV$ & $10^{12}\GeV$ \\  \colrule
		$10^{12}\GeV$ & $3.16\times 10^{12}\GeV$ & $10^{13}\GeV$ \\ \botrule
\end{tabular}
\caption{\label{tab:MassRH} 
RH neutrino masses used to obtain the baryon-to-photon ratio, see main text.
 }
\end{table}

\section{D. Gravitational waves from PBHs}

\begin{figure}[t!]
\includegraphics[width=0.45\textwidth]{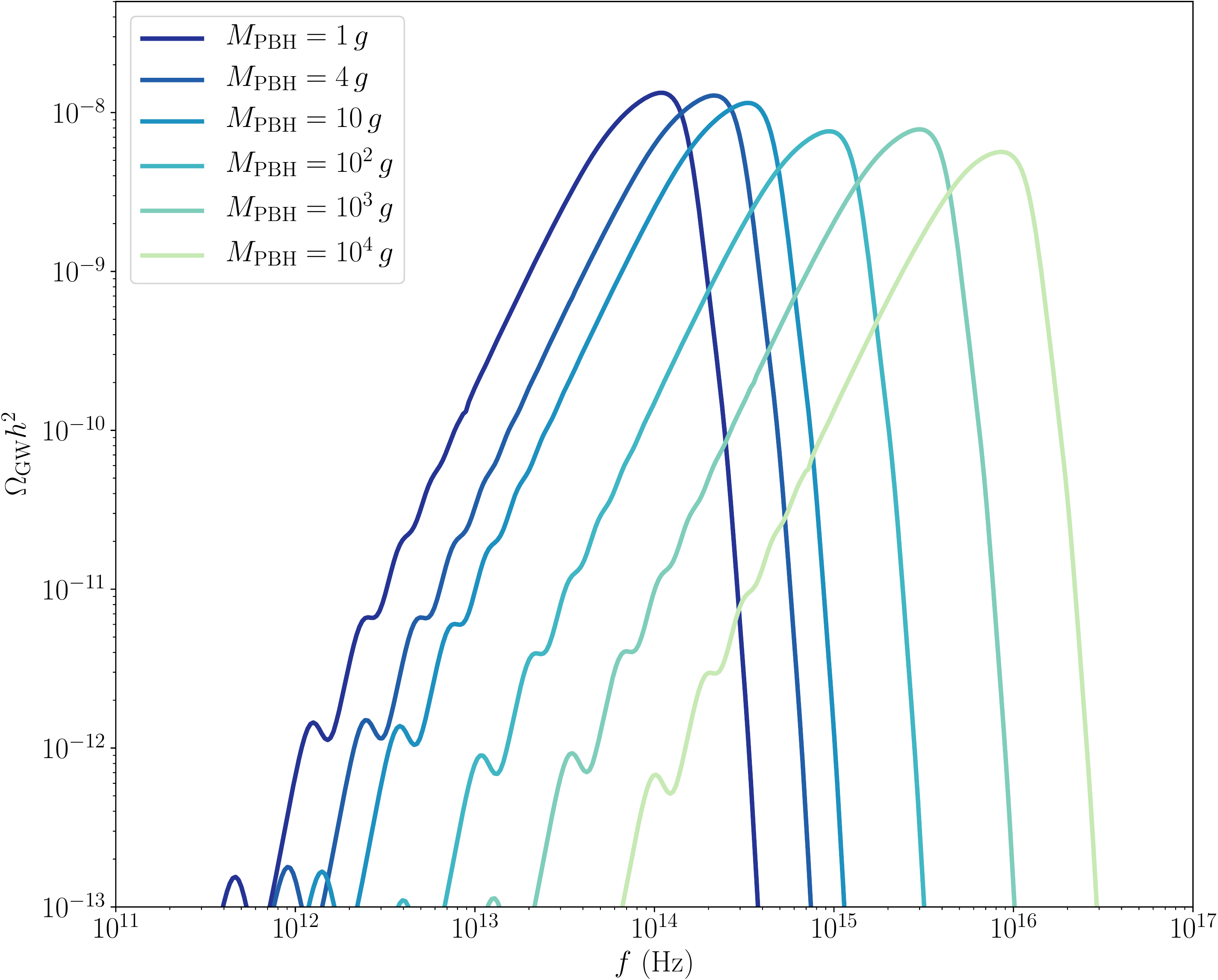}
\caption{Stochastic gravitational wave spectrum generated by evaporating primordial black holes.}\label{fig:GW}
\end{figure}

As an example, let us examine the stochastic background of GWs produced directly from the PBHs evaporation. Their energy density can be obtained by integrating the Hawking spectrum over the PBH lifetime,
\begin{align}
\Omega_{\rm GW}h^2 (f_0) = \frac{8(2\pi)^4f_0^4}{3H_0^2M_{\rm Pl}^6}\frac{\varrho_{\rm PBH}^i}{M_i}\int_{1}^{a_{\rm EV}}\frac{da}{a^4 H}\frac{\sigma^2_{\rm abs}(2\pi G M a_0 f_0/ a) M^2}{\exp\left[16\pi^2 G M a_0 f_0/a\right]-1}\,,
\end{align}
being $M_{\rm Pl} = G^{-1/2}$ the Planck mass, $\frac{d^2 {\cal N}_g}{dp\,dt}$ the Hawking emission rate for gravitons, $f_0$ the GW frequency and $a_0$ the scale factor today. Employing the solutions of Eqs.~\eqref{eq:BEnolog} and \eqref{eq:BELeplog}, we present the GW spectrum for a range of PBH masses in  \figref{fig:GW} which are relevant for thermal leptogenesis. The peak frequency of the GW spectrum generated from evaporating PBHs is larger for higher initial masses. This shift occurs because the redshift effects are stronger for GW emitted from lighter PBHs, which evaporated earlier in the history of the universe. 
%
%

%\bibliographystyle{apsrev4-1}
%\bibliography{ref}{}

\end{document}